\documentclass[%
 reprint,
superscriptaddress,
 amsmath,amssymb,
 aps,
]{revtex4-2}
\usepackage[normalem]{ulem}
\usepackage{upgreek}
\usepackage{xcolor}
\usepackage{listings}
\usepackage{caption}
\lstdefinestyle{Mathematica}{
    language=Mathematica,
    basicstyle=\ttfamily\small,
    commentstyle=\color{green!40!black},
    keywordstyle=\color{black},
    stringstyle=\color{purple},
    numbers=left,
    numberstyle=\tiny\color{gray},
    numbersep=5pt,
    breaklines=true,
    frame=single,
    showstringspaces=false,
    tabsize=2,
    columns=fullflexible,
    mathescape=true,
    moredelim=[s][\color{purple}]{@}{@} 
}
\newcommand{\norm}[1]{\left\lVert#1\right\rVert}
\usepackage{graphicx}
\usepackage{dcolumn}
\usepackage{bm}

\definecolor{grey}{RGB}{140,140,140}
\graphicspath{{./},{./figures/}}
\begin{document}

\preprint{APS/123-QED}

\title{Refined bounds on energy harvesting from 
anisotropic fluctuations}

\author{Jordi Ventura Siches}
 
\author{Olga Movilla Miangolarra}%
 
\author{Tryphon T. Georgiou}
\affiliation{%
 Department of Mechanical and Aerospace Engineering,\\ University of California, Irvine, California 92697, USA}%

\date{\today}

\begin{abstract}
We consider overdamped Brownian particles with two degrees of freedom (DoF) that are confined in a time-varying quadratic potential and are in simultaneous contact with heat baths of different temperatures along the respective DoF.
The anisotropy in thermal fluctuations can be used to extract work by suitably manipulating the confining potential.
The question of what the maximal amount of work that can be extracted is has been raised in recent work, and has been computed under the simplifying assumption that the entropy of the distribution of particles (thermodynamic states) remains constant throughout a thermodynamic cycle.
Indeed, it was shown that the maximal amount of work that can be extracted amounts to solving an isoperimetric problem, where the $2$-Wasserstein length traversed by thermodynamic states quantifies dissipation that can be traded off against an area integral that quantifies work drawn out of the thermal anisotropy.
Here, we remove the simplifying assumption on constancy of entropy. 
We show that the work drawn can be computed similarly to the case where the entropy is kept constant while the dissipation can be reduced by suitably tilting the thermodynamic cycle in a thermodynamic space with one additional dimension. Optimal cycles can be locally approximated by solutions to an isoperimetric problem in a tilted lower-dimensional subspace.
\footnote{Supported by the ARO under grant W911NF-22-1-0292 and the AFOSR under grant FA9550-23-1-0096. OMM was supported by ``la Caixa'' Foundation (ID 100010434) with code LCF/BQ/AA20/11820047.\\ {Emails: \{jordiv, omovilla, tryphon\}@uci.edu}}
\end{abstract}

\maketitle


\section{Introduction}
In recent years the newly developed field of Stochastic Thermodynamics has made it possible to quantify energy exchanges between thermodynamic systems taking place in finite time and has provided models for studying naturally occurring processes transducing energy at a cellular level \cite{sekimoto,seifert_stochastic,peliti}. 
In this endeavor, a paradigmatic example that allows harvesting mechanical work from thermal gradients is the Brownian gyrator \cite{brownian}--a simple model that
can sustain a far-from-equilibrium operation powered by anisotropic thermal excitations. 
Detailed analysis of a thermodynamic cycle was first carried out in 
\cite{olga} and subsequently in \cite{geom} to characterize optimality and to derive achievable bounds for power and efficiency that a so-enacted thermodynamic engine is capable of. The analysis of the respective dynamical process in \cite{olga} was carried out under a simplifying assumption that the entropy of the system-states remains constant. Here, we remove this restriction and show how the conclusions in \cite{olga} extend to thermodynamic cycles traversing more general system states. Specifically, 
we conclude that work is harvested from the two heat baths in a similar manner as in \cite{olga}, but the flexibility of traversing paths corresponding to states of different entropy allows a reduction in the dissipation, thereby increasing the net work being extracted and attaining higher efficiency.

\section{Model}
The Brownian gyrator \cite{brownian} represents a thermodynamic system of overdamped particles having two coupled degrees of freedom (DoF) that are subject to thermal excitation at different temperatures, $T_1$ and $T_2$.
These two DoF may represent the position $X_t\in\mathbb R^2$  of a particle on the plane at time $t$, with $\{X_t\mid t\in\mathbb R\}$ a stochastic process obeying the (overdamped) Langevin dynamics
\begin{equation*}
    dX_t = -\gamma^{-1}\nabla U(t,X_t) dt + \sqrt{\frac{2k_BT}{\gamma}}\, dB_t.
\end{equation*}
Here, $\{B_t\mid t\in\mathbb R\}$ is a two-dimensional Brownian motion, $T=\text{diag}(T_1,T_2)$ is a diagonal matrix of the two temperatures, $\gamma$ is the dissipation constant, and $U(t,X_t)$ is a time-varying potential. 

The state of the Brownian gyrator is the distribution $\rho(t,x)$ of the Langevin particles, with $x\in\mathbb R^2$, which obeys the Fokker-Planck equation $\partial_t \rho + \nabla\cdot J = 0$ for the probability current $J=-\rho (\nabla U +k_B T\nabla\log(\rho))/\gamma$.
The problem we consider is to steer $\rho(t,x)$ along a closed orbit (thermodynamic cycle) via suitable manipulation of the controlling potential $U(t,X_t)$ so as to maximize work extracted from the coupling of the system with heat baths for specified dissipation.
The mechanical power exchanged via manipulating $U$ can be expressed
as $\dot {\mathcal W}=\int_{\mathbb R^2}\rho\partial_t U dx$ and the rate of heat uptake from the two reservoirs as $\dot {\mathcal Q}=-\int_{\mathbb R^2}U\nabla\cdot J dx$, see
\cite[page 212]{sekimoto}.

We specialize to the case of a quadratic potential $U(t,X_t) = \tfrac{1}{2}\, X_t^TK(t)X_t$ centered at the origin. Actuation is effected via suitable schedule for the time-varying ``spring-matrix'' $K(t)$. The thermodynamic state $\rho$ remains a two-dimensional Gaussian distribution with mean equal to zero and covariance matrix $\Sigma(t)\in\mathbb R^{2\times 2}$ that
satisfies
\begin{equation}\label{eq:lyap}
    \gamma\dot{\Sigma}(t) = -K(t)\Sigma(t) - \Sigma(t)K(t)+2k_BT.
\end{equation}
Rates of work and heat exchange between the system, the actuating potential and the two heat baths can be readily expressed in terms of $K(t)$ and $\Sigma (t)$. Indeed,
the internal energy of the system is
\begin{equation*}
    \mathcal{E} 
    = \tfrac{1}{2} \text{Tr}[K(t)\Sigma(t)],
\end{equation*}
where $\text{Tr}[\cdot]$ denotes the trace operator. Work and heat-exchange rates are given by (see \cite{olga}, \cite[page 212-213]{sekimoto})
\begin{align*}
    \dot{\mathcal{W}}(t) = \tfrac{1}{2} \,\text{Tr}[\dot{K}(t)\Sigma(t)],\mbox{ and }  \dot{\mathcal{Q}}(t)  = \tfrac{1}{2} \,\text{Tr}[K(t)\dot{\Sigma}(t)].
\end{align*}

The ``spring matrix'' $K(t)$ can be expressed from \eqref{eq:lyap} as a function of $(\Sigma(t),\dot{\Sigma}(t))$,
\begin{equation*}
\begin{split}
    K(t) &= \int_0^\infty e^{-\tau \Sigma(t)} \left(2k_B T-\gamma \dot{\Sigma}(t)\right)e^{-\tau \Sigma (t) }\, d\tau\\
    &= : \mathcal{L}_{\Sigma(t)}[2k_B T-\gamma \dot{\Sigma}(t)].
\end{split}
\end{equation*}
Substituting this expression for $K(t)$ into the formula for $\dot{\mathcal Q}$, the heat-exchange rate splits into two terms, one that is linear in $\dot \Sigma$ and one that is quadratic,
\begin{align*}
    \dot{\mathcal{Q}}=k_B \text{Tr}[\mathcal{L}_{\Sigma(t)}[T]\dot{\Sigma}(t)] -\frac{\gamma}{2} \text{Tr} [\mathcal{L}_{\Sigma(t)}[\dot{\Sigma}(t)]\dot{\Sigma} (t)].
\end{align*}
The linear term represents quasi-static heat, as it remains invariant with the speed of traversing the path, while the quadratic quantifies dissipation as it vanishes when the speed slows down to $0$.
Thus, the total quasi-static heat and dissipation over a cycle with period $t_f$ are
\begin{equation}\label{eq:defQ}
\begin{split}
    \mathcal{Q}_\text{qs}&= k_B \int_0^{t_f} \text{Tr}\left[\mathcal{L}_{\Sigma(t)}[T]\dot{\Sigma}(t)\right]dt, \mbox{ and}\\
    \mathcal{Q}_\text{diss} &= \frac{\gamma}{2} \int_0^{t_f} \text{Tr}\left[\mathcal{L}_{\Sigma(t)}[\dot{\Sigma}(t)]\dot{\Sigma}(t)\right] dt,
    \end{split}
\end{equation}
respectively \cite{olga}. 

The state $\rho$ of the thermodynamic system at time $t$, being zero-mean Gaussian, is specified by its covariance $\Sigma(t)$. Thus, we seek to study thermodynamic cycles as closed orbits on the {\em space of positive definite $2\times 2$ real symmetric matrices}, the $\Sigma$-space. To this end, we select coordinates $(r,\theta,z)$ for this space as
\begin{align*}
    r&=\frac12 \log(\lambda_1(\Sigma)/\lambda_2(\Sigma))\\
        z&=\log(\det(\Sigma)) \;\;= \log(\lambda_1(\Sigma)\cdot\lambda_2(\Sigma)),
\end{align*} 
where  $\lambda_{1,2}$ denote the eigenvalues of $\Sigma$, $\lambda_1\geq \lambda_2>0$, and $\theta$ specifies the rotation matrix
\begin{equation*}
    R\left(-\tfrac{\theta}{2}\right) = \begin{pmatrix}
        \cos\left(\tfrac{\theta}{2}\right) & -\sin\left(\tfrac{\theta}{2}\right)\\[.02in]
        \sin\left(\tfrac{\theta}{2}\right) & \phantom{-}\cos\left(\tfrac{\theta}{2}\right)
    \end{pmatrix}
\end{equation*}
\begin{figure}[h]
    \centering   \includegraphics[width=0.45\textwidth]{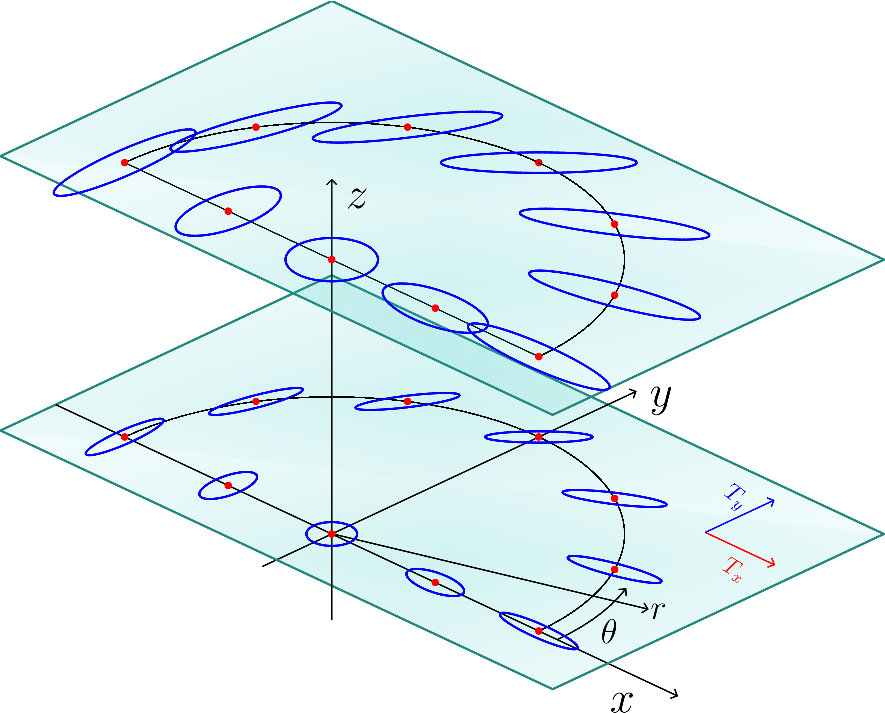}
    \caption{Pictorial of planar closed orbits on the $\Sigma$-space.}
    \label{fig:parametrization}
\end{figure}
that diagonalizes $\Sigma$. Thus,
\begin{equation}\label{eq:eq2}
    \Sigma = R\left(-\tfrac{\theta}{2}\right) \sigma(z,r)R\left(-\tfrac{\theta}{2}\right)^\prime,
\end{equation}
where $\,^\prime$ denotes transposition, and
\begin{equation*}
    \sigma (z,r) = e^{\tfrac{z}{2}}\begin{pmatrix}
        e^{r} & 0 \\ 0 & e^{-r}
    \end{pmatrix}=\begin{pmatrix}
        \lambda_1 & 0 \\ 0 & \lambda_2
    \end{pmatrix}.
\end{equation*}

In the $\Sigma$-space, $(r,\theta)$ are planar polar coordinates and specify the eccentricity and orientation of principle components of $\Sigma$, respectively, while $z$ relates to the area of drawn ellipses and specifies the entropy ($-\int_{\mathbb R^2}\rho\log\rho dx=\frac12 z + \rm const.$) of $\rho$.
Figure \ref{fig:parametrization} displays closed semi-circular planar orbits in which $z$ is kept constant.
The analysis in \cite{olga} was carried out for $z$ constant. In the sequel we will explore the case where $z$ is not kept constant and seek properties of optimizing cycles.

We are interested in closed orbits that maximize work produced, namely, closed paths in the $\Sigma$-space solving
\begin{align}\label{eq:problem}
    \max_{\Sigma (t)}\,  
    & \, \, \, \mathcal{Q}_\text{qs} (\Sigma(t) )  - \mathcal{Q}_\text{diss} (\Sigma (t))
\end{align}
subject to $\Sigma (0)=\Sigma (t_f)$.
We also define efficiency \textcolor{black}{\cite{brandner}}
\begin{equation}\label{eq:def_eff}
    \eta := \frac{\mathcal{W}_{\rm out}}{\mathcal{Q}_\text{qs }} = 1 - \frac{\mathcal{Q}_{\rm diss}}{\mathcal{Q}_{\rm qs}}
\end{equation}
as the ratio between the work output $\mathcal{W}_{\rm out}$ over a cycle 
\begin{equation*}
    \mathcal{W}_{\rm out} = \mathcal{Q}_{\rm qs} - \mathcal{Q}_{\rm diss} 
\end{equation*}
and the work output in the quasi-static limit $\mathcal{Q}_{\rm qs}$. \textcolor{black}{Definition \eqref{eq:def_eff} differs from the more traditional one where work is compared to heat drawn from a hot heat bath, and captures the dissipation along the cycle. 
Trivially $\eta\leq 1$, with equality attained as $t_f\to\infty$.}

In the following section we explain how \eqref{eq:problem} relates to an isoperimetric problem, seeking a maximal area-integral for a fixed ($2$-Wasserstein) length traversed in the $\Sigma$-space, echoing results in \cite{olga} and  \cite{frim} for the cases of constant $z$ and linear response regime, respectively. We also provide a correction to the bound $\eta\leq 1$ that takes into account the finite period in traversing thermodynamic cycles, and compute efficiency at maximum power.

\section{Results and analysis}

Taking the time derivative of $\Sigma$ in \eqref{eq:eq2} we obtain 
\begin{equation}\label{eq:eq3}
    \dot{\Sigma} = \tfrac{1}{2} \, R(\sigma \dot{z} + 2\sigma \Xi \dot{r} + ( \sigma \Omega -\Omega \sigma )\dot{\theta}) R^\prime,
\end{equation}
for
\begin{equation*}
    \Xi = \begin{pmatrix}
        1 & 0 \\ 0 & -1
    \end{pmatrix}, \quad \text{and}\quad \Omega =\begin{pmatrix}
        0 & 1 \\-1 & 0
    \end{pmatrix} .
\end{equation*}
To tackle \eqref{eq:problem}, we express the quasi-static heat and dissipation over a cycle 
\eqref{eq:defQ} explicitly in terms of parameters $(r,\theta,z)$ using \eqref{eq:eq3}. This is as follows,
\begin{equation}\label{eq:Qall}
\begin{split}
\mathcal{Q}_\text{qs} =& \frac{k_B\Delta T}{2}\int_0^{t_f}
\cos(\theta)\dot{r}-\tanh(r)\sin(\theta)\dot{\theta}\,dt\\
\mathcal{Q}_\text{diss} =& \frac{\gamma}{2}\int_0^{t_f}e^{z/2} \Big(\sinh\left(r\right)\tanh\left(r\right)\dot{\theta}^2\\&\ \quad+\cosh\left(r\right)\left(\dot{r}^2+\tfrac{1}{4}\dot{z}^2\right)+\sinh\left(r\right)\dot{r}\dot{z}\Big)dt,
\end{split}
\end{equation}
for $\Delta T=T_1-T_2$. (Compare with \cite[Equations (7a-7b)]{olga}, where $z=\rm \,const.$ and thus $\dot{z}=0$). 
Interestingly, varying $z$ (the entropy of thermodynamic states) along cycles does not affect $\mathcal{Q}_{\rm qs}$ but impacts dissipation $\mathcal{Q}_{\rm diss}$.    
Thus, by suitably modifying the state-entropy along a thermodynamic cycle, trajectories of a fixed length (i.e., fixed dissipation) can encompass a larger area in the $\Sigma$-space and thereby enable a relative increase in work production. Symbolic code that can be used to obtain Eq.~(\ref{eq:Qall}) is given in the Appendix \ref{section:app2}.

\subsection{Geometric analysis}
We now explain the inherently geometric nature of the problem. Firstly, dissipation can be expressed as
\begin{equation}\label{eq:Qdiss}
    \mathcal{Q}_\text{diss} = \frac{\gamma}{2}\int_0^{t_f} \norm{\dot{\alpha}(t)}^2_{g} \, dt,
\end{equation}
where $\alpha(t) = \{(r(t),\theta(t), z(t)):\, t\in[0,t_f] \}$ denotes a trajectory on the $\Sigma$-space and $\norm{\cdot}^2_{g}$ denotes the square norm of a vector with respect to the metric
\begin{equation*}
    g = e^{z/2} \begin{pmatrix}
        \cosh(r) & 0 & \tfrac{1}{2}\sinh(r)\\
        0 & \sinh(r)\tanh(r) & 0 \\
        \tfrac{1}{2}\sinh(r) & 0 &\tfrac{1}{4}\cosh(r)
    \end{pmatrix}.
\end{equation*}
Note that when $z$ remains constant, the metric in~\cite{olga} is recovered. By the Cauchy-Schwarz inequality, the minimal dissipative heat for any trajectory is given by
\begin{equation}\label{eq:dontgiveup}
  \frac{\gamma}{2t_f} \left(\int_0^{t_f}\norm{\dot{\alpha}(t)}_g \, dt\right)^2 = : \frac{\gamma}{2t_f} \, \ell^2 
\end{equation}
where $\ell$ is the length of the path with respect to $g$ on the $\Sigma$-space, which coincides with $\mathcal Q_{\rm diss}$, and hence, with the $2$-Wasserstein length of the cycle of thermodynamic states \cite{aurell2011optimal}.
The minimum is attained when $\norm{\dot{\alpha}(t)}_g$ remains constant along the path.

Second, the quasi-static heat can be written as a weighted surface integral over a domain $\mathcal D$, precisely as shown in \cite{olga}. Note however, that $\mathcal{D}$ is no longer the domain encircled by a path drawn on some two-dimensional submanifold in the $\Sigma$-space, but instead, it is the area enclosed by the projection of the cycle 
onto a plane that corresponds to a constant value for $z$.

Following \cite{olga}, by means of Stokes' theorem, quasi-static heat from \eqref{eq:Qall} gives
\begin{align}\label{eq:QqsA}
    \mathcal{Q}_\text{qs} &= \pm \frac{k_B\Delta T}{2}\iint_\mathcal{D} \frac{\tanh^2(r)}{r}\, \sin\theta \,rd\theta\,dr\\\nonumber
    &=:\pm \frac{k_B\Delta T}{2}\, \mathcal{A}_h,
\end{align}
where the $\pm$ sign depends on the direction chosen, and
\begin{equation*}
    \mathcal{A}_h = \iint_\mathcal{D} h(r,\theta, z) \sqrt{\text{det}(g)}\, dr d\theta,
\end{equation*}
is an area integral with respect to the Riemannian canonical $2$-form of the metric $g$ and the work density function
\begin{equation}\label{eq:workdensity}
    h(r,\theta, z) = 2e^{-3z/4} \sin\theta \, \frac{\tanh(r)}{\sqrt{\cosh(r)}},
\end{equation}
that results in $\mathcal{A}_h$ being independent of $z$.

Thus, the problem to maximize work extraction $\mathcal{W}_\text{out}$ along a cycle, namely,
\begin{equation}\label{eq:isop}
    \max_{\alpha(t)}  \, \, \mathcal{A}_h - \mu \ell^2
\end{equation}
for a given $\mu = \frac{\gamma}{k_B\Delta T t_f}$, that can be interpreted as a Lagrange multiplier, amounts to maximizing the area integral $\mathcal{A}_h$ enclosed by a cycle of fixed length $\ell$, as in \cite{olga}. {\color{black} Since $\mu$ acts as a penalty on the length of the cycle, it is clear that larger values of $\mu$ lead to smaller values for the optimal length $\ell$.}
On the other hand, efficiency (see \eqref{eq:def_eff}) can also be expressed in geometric terms as 
\begin{equation*}
    \eta = 1 - \mu\frac{\ell^2}{\mathcal{A}_h},
\end{equation*}
with the problem to maximize efficiency along a cycle of arbitrary length turning into a search for an isoperimetric-like inequality in the space of thermodynamics states. That is, maximizing efficiency amounts to seeking
$$
\mu^*:=\max_\mathcal D\frac{\mathcal A_h}{\ell^2}.
$$


\textcolor{black}{We remark that, due to the exponential term $e^{z/2}$ in the expression for the metric $g$, increasingly negative values of $z$ (and thus, increasingly negative entropy) result in a vanishingly small dissipation. 
However, such a tight confinement requires an arbitrarily strong potential. To ensure physically meaningful conditions, we specify a starting value for $z$ along the cycle, which amounts to specifying the entropy at that point.
}




\subsection{Local analysis}
{\color{black} To gain intuition on the shape of the optimal curves, we perform a local analysis, valid when $\ell\to 0$. 
To this end, we consider cycles of infinitesimally small length around an operating point $(r_0,\theta_0,z_0),$ which are optimal for large enough values of the parameter $\mu$.
}

Since the work density \eqref{eq:workdensity} is proportional to $\sin(\theta)$ and  $g$ is independent of $\theta$, the choice $\theta_0 = \tfrac{\pi}{2}$ maximizes $\mathcal A_h$
locally. 
Without loss of generality we also fix $z_0=0$ (corresponding to $\det(\Sigma) = 1$) and consider closed paths $\{( r(t), \theta(t), z(t)\mid t\in[0,t_f]\}$ with
\begin{align*}
        r(t) & = r_0 + \epsilon r_1 (t)\\
         \theta (t) &= \frac{\pi}{2} + \epsilon\theta_1(t)\\
        z (t) &= \epsilon z_1(t),
\end{align*}
for $r_0$ to be determined with $\epsilon>0$ assumed small.
Up to $o(\epsilon^2)$, the quasi-static and dissipative heat are
\begin{align}\label{eq:Qepsilon}
\begin{split}
  \mathcal{Q}^\epsilon_\text{qs}  &= -\frac{k_B\Delta T}{2} \,\epsilon^2\int_0^{t_f} \theta_1\dot{r}_1 + \frac{1}{c_0^2}\, r_1\dot{\theta}_1\, dt\\
\mathcal{Q}^\epsilon_\text{diss} \! &= \! \frac{\gamma}{2} \epsilon^2\!\! \int_0^{t_f} \!\!\!\left(c_0\dot{r}_1^2 + \frac{s_0^2}{c_0} \dot{\theta}_1^2 + \frac{1}{4}c_0\dot{z}_1^2 + s_0\dot{r}_1\dot{z}_1\!\right)\! dt,
\end{split}
\end{align}
where $c_0 = \cosh(r_0)$ and $s_0 = \sinh(r_0)$. Note that $\mathcal Q^\epsilon_{\rm qs}$ can be written as an area integral as before, or as the line integral above.

We now consider maximizing work extracted, namely,
\begin{equation}\label{eq:problem-linearized}
    \max_{ r, \theta, z}\  \mathcal{Q}^\epsilon_\text{qs}-\mathcal{Q}^\epsilon_\text{diss},
\end{equation}
over a small cycle about $(r_0,\theta_0=\frac{\pi}{2},z_0=0)$.
The Euler-Lagrange equation (first order necessary condition for optimality) for the functional in \eqref{eq:problem-linearized} with respect to the $z$-coordinate gives
\begin{equation}\label{eq:plane}
    \ddot{z}_1 = -2\,\frac{s_0}{c_0}\, \ddot{r}_1 \;\;(= -2\tanh(r_0)\ddot{r}_1).
\end{equation}
Integrating over time gives $\dot z_1 =-2\frac{s_0}{c_0}\dot r_1 +\rm const.$ This constant however must be equal to zero on periodic orbits.
Substituting $\dot z_1 =-2\frac{s_0}{c_0}\dot r_1$ into \eqref{eq:Qepsilon}, 
we obtain that
\begin{equation}\label{eq:Qdiss-noz}
 \mathcal Q^\epsilon_{\rm diss} = \frac{\gamma \epsilon^2}{2}\int_0^{t_f}\|(\dot r_1,\dot\theta_1)\|^2_{g_0}dt   
\end{equation}
is quadratic in the $2$-dimensional velocity vector $(\dot r,\dot\theta)$ for the metric (cf.\ \eqref{eq:Qdiss})
\begin{align*}
    g_0 = \begin{pmatrix}
        \frac{1}{c_0} & 0 \\ 0 & \frac{s_0^2}{c_0}
    \end{pmatrix}.
\end{align*}

As observed in the geometric analysis section, instead of Problem \eqref{eq:problem-linearized} we may instead consider the problem to maximize
$\mathcal Q^\epsilon_{\rm qs}$ subject to
$\mathcal Q^\epsilon_{\rm diss}$ in \eqref{eq:Qdiss-noz} being specified. To this end, cf.\ \eqref{eq:dontgiveup}, we set
\begin{equation}\label{eq:constraint}
   \frac{\mathcal Q^\epsilon_{\rm diss}}{\epsilon^2} = \frac{\gamma}{2}\int_0^{t_f} \frac{1}{c_0}\dot r_1^2 + \frac{s_0^2}{c_0}\dot\theta_1^2 \, dt =: \frac{\gamma\ell^2_{\epsilon}}{2t_f}.
\end{equation}

Our problem now becomes
\begin{equation*}
    \max_{r_1(t),\theta_1(t)} \int_0^{t_f}\!\!\underbrace{\left[ \theta_1 \dot r_1 + \frac{1}{c_0^2} r_1\dot\theta_1 \!+\!\lambda \left(\frac{1}{c_0}\dot r_1^2 + \frac{s_0^2}{c_0}\dot\theta_1^2 - \frac{\ell^2_{\epsilon}}{t_f^2}\right) \!\right]}_{\mathcal{L}(r_1,\theta_1,\dot r_1,\dot\theta_1)}\! dt,
\end{equation*}
where the constraint \eqref{eq:constraint} has been included in the Lagrangian $\mathcal{L}$ with Lagrange multiplier $\lambda$. The corresponding Euler-Lagrange equations are
\begin{equation*}
    \ddot r_1 = -\frac{ s_0^2}{2\lambda c_0}\dot \theta_1,\qquad \ddot \theta_1 = \frac{1}{2\lambda c_0}\dot r_1,
\end{equation*}
giving
\begin{equation*}
    \begin{split}
        r_1(t) &= s_0 A \cos(\omega t) - s_0 B\sin(\omega t)\\
        \theta_1(t) &=\phantom{s_0} A \sin(\omega t) + \phantom{s_0}B\cos(\omega t),
    \end{split}
\end{equation*}
with $\omega = \tfrac{2\pi}{t_f} = \tfrac{s_0}{2\lambda c_0}$.
Since we are interested in the complete (closed) orbit, we may take $B=0$. Then, $A =\tfrac{\ell_{\epsilon}\sqrt{c_0}}{2\pi s_0}$ from \eqref{eq:constraint}, and we obtain the equations of an ellipse
\begin{equation}\label{eq:local_solution}
    \begin{split}
        r_1 (t) &= \frac{\ell_{\epsilon} \sqrt{c_0}}{2\pi} \, \cos \left(\tfrac{2\pi}{t_f} t\right)\\
        \theta_1 (t) & = \frac{\ell_{\epsilon} \sqrt{c_0}}{2\pi s_0} \, \sin \left(\tfrac{2\pi}{t_f} t \right),
    \end{split}
\end{equation}
for $t\in[0,t_f]$. 

Up to $\epsilon^2$, the quasi-static heat and dissipation are
\begin{align*}
    \mathcal{Q}^\epsilon_\text{qs} = \frac{k_B\Delta T s_0\ell^2_{\epsilon} }{8\pi c_0}\, \epsilon^2,
    \mbox{ and } \mathcal{Q}^\epsilon_\text{diss} = \frac{\gamma\ell^2_{\epsilon}  }{2 t_f}\, \epsilon^2,
\end{align*}
respectively, giving
\begin{figure}[t]
    \centering
    \includegraphics[width=0.45 \textwidth]{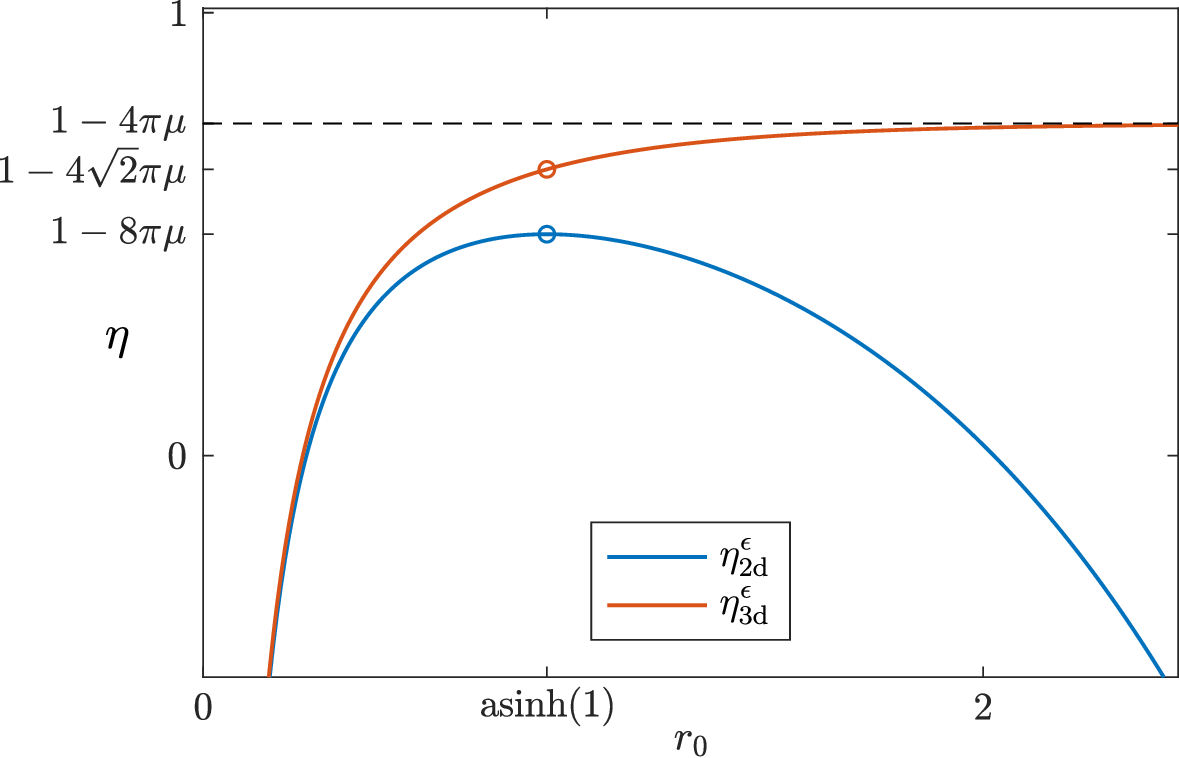}
    \caption{Efficiency of isentropic (blue, bottom) and not-isentropic/general (red, top) cycles vs.\ $r_0$ (eccentricity of states);
    locally for $\mu=\tfrac{1}{16\pi}$, $\theta_0=\tfrac{\pi}{2}$, $z\simeq 0$. 
    }
    \label{fig:efficiency}
\end{figure}
$$
\frac{1}{\mu}\frac{\mathcal{Q}_{\rm qs}^\epsilon}{\mathcal{Q}^\epsilon_{\rm diss}}= \frac{1}{4\pi \mu}\frac{s_0}{c_0}
$$
Thus, the efficiency can be expressed as
\begin{equation}\label{eq:efficiency}
    \eta^\epsilon_{3\rm d} = 1 - \frac{\mathcal{Q}^\epsilon_{\rm diss}}{\mathcal{Q}_{\rm qs}^\epsilon} 
    = 1- 4\pi \mu \frac{c_0}{s_0}.
\end{equation}
Here, the subscript $3\rm d$ refers to the fact that optimization takes place in the $3$-dimensional parameter space, of coordinates $(r,\theta,z)$. The efficiency is seen to be greater than that obtained over cycles of constant entropy ($z=\rm constant$), when it was found that   $\eta^\epsilon_{2\rm d} = 1 - 4\pi \mu \, \frac{c_0^2}{s_0}$\cite{olga}. It was further shown in \cite{olga} that $\eta_{2\rm d}^\epsilon=1-8\pi\mu$ is maximal (achieved for $r_0 = \text{asinh}(1)$). Instead, $\eta^\epsilon_{3\rm d}$ increases as $r_0\to\infty$ towards the limit $1-4\pi\mu$,  see Figure \ref{fig:efficiency}.

Equation \eqref{eq:efficiency} implies an inherent speed limit, in that for 
\begin{equation}
 t_f < \frac{4\pi \gamma c_0}{s_0 k_B \Delta T}    \label{eq:sl}
\end{equation}
it is impossible to extract work ($\eta_{3\rm d}^\epsilon <0$). A corresponding speed limit for isentropic cycles obtained in \cite{olga}, $t_f < \tfrac{4\pi \gamma c^2_0}{s_0k_B\Delta T}$, differs by a factor of $c_0$.

For isentropic cycles, it was inferred in \cite{olga} that the efficiency - and thus the ratio of the weighted area and the $2$-Wasserstein length of the cycle squared - was maximized as $\ell\to 0$. 
In the present case of cycles where the entropy is allowed to vary, it appears that the same is true, and thus we presume the efficiency to be bounded by $\eta\,\leq 1-4\pi\, \frac{t_c}{t_f}$, where $t_c = \mu t_f$ is a characteristic time, pointing towards a more general isoperimetric inequality and speed limit. Numerical experiments appear to back this hypothesis but a formal proof is lacking.

\textcolor{black}{Both limits  $\ell\to 0$ (with $t_f$ fixed) and $t_f\to \infty$ (with $\ell$ fixed) represent quasi-static operation, for which the cycle is traversed arbitrarily slowly, leading to vanishing dissipation. 
Remarkably, these two scenarios are distinctly different. The limit of arbitrarily slow operation is achieved by lengthening the time to complete the cycle in one case, and by shrinking the path to be traversed in the other. 
This makes the efficiency different in the two limits. Specifically, as $t_f\to\infty$, $\eta\to 1$, as the process becomes quasi-static in the traditional sense.
In the other case, as $\ell\to 0$ (with a finite $t_f$), dissipation vanishes at the same rate as the quasi-static work, leading to a negative contribution to the efficiency and $\eta\to 1-8\pi\mu<1.$}



\subsection{General cycles}
We now consider {\color{black}general cycles of finite lengths $\ell$, that are in correspondence with values of $\mu$.} Since the Riemannian metric $g$ can no longer be assumed constant, closed-form expressions for the optimal cycles are not feasible\textcolor{black}{, and the cycles are constructed numerically. Specifically, we solve \eqref{eq:isop} numerically by fixing the length $\ell$ of the cycle, interpreting $\mu$ as a Lagrange multiplier. 
To this end, we implement gradient descent on the space of functions $(r(t),\theta(t),z(t))$ to determine cycles that maximize work extraction for different lengths.
}


\begin{figure}[tb]
    \centering
\includegraphics[width=0.45\textwidth]{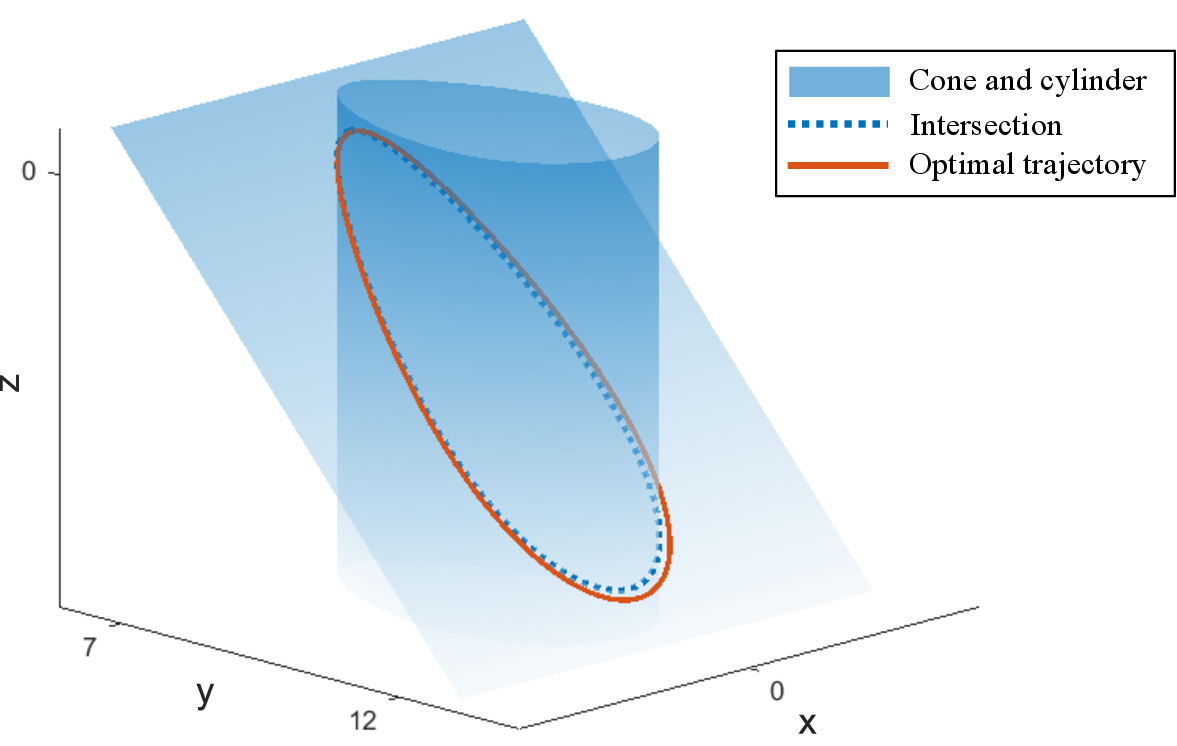}
    \caption{Optimal cycle (red, solid) and its approximation (blue, dotted) as the intersection of a cone with an elliptic cylinder for $\ell=0.1$}
    \label{fig:cone_cylinder}
\end{figure}

\begin{figure}[t]
    \hspace*{-10pt}\includegraphics[width=0.5\textwidth]{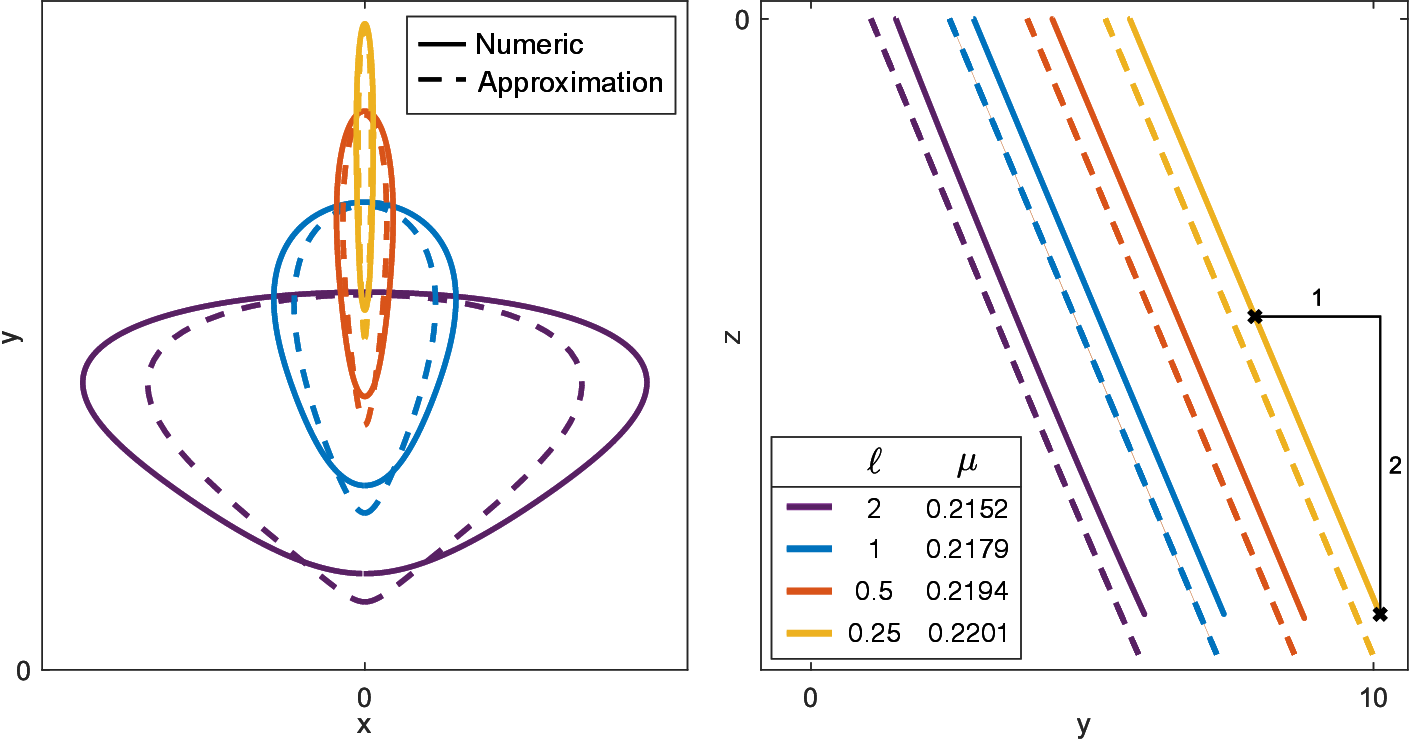}
    \caption{Optimal trajectories of different length and their local approximations as seen from above, along the $z$-axis (left), and from the side, along the $x$-axis (right).}
    \label{fig:all_cycles}
\end{figure}

{\color{black}The discrepancy
between a numerically obtained optimal cycle and its local approximation is negligible for small $\ell$ (corresponding to large $\mu$),} as highlighted in Figure \ref{fig:cone_cylinder}.
Note that the coordinates $x=r\cos \theta$ and $y = r\sin\theta$ have been used as axes (in accordance with their portrayal in Figure~\ref{fig:parametrization}).
As depicted in the figure, the approximation is precisely the intersection of a cone with an elliptic cylinder \eqref{eq:local_solution}; 
the equation of the cone 
$z_1=-2\tanh(r_0)r_1+\,\rm const.$ follows from \eqref{eq:plane}.
Due to the scale of the figure (range of $y$-values far from the origin), the slice of the cone appears as planar.

For larger values of $\ell$, optimal cycles are depicted in Figure \ref{fig:all_cycles}.  The dashed curves on the left plot outline the local approximations of optimal cycles. These are in surprisingly good agreement with the exact numerical solutions (solid curves).
The subfigure on the right displays the side-view of optimal cycles, lying on the generatrix of the cone with slope $-2 \tanh(r_0)\simeq -2$ (since here, $r_0$ is large).


\subsection{Efficiency at maximum power}

The thermodynamic efficiency of heat engines is maximal in the quasi-static limit ($t_f\to \infty$), a regime with vanishing power output.
Indeed, quantifying the power that an engine is capable of has motivated this and earlier studies. In the regime where power is maximal, it is also of interest to quantify
efficiency.

In the present context, for a specified thermodynamic cycle (hence, with $\mathcal{A}_h,\ell$ given), the power output 
\begin{equation*}
    P = \frac{\mathcal{W}_{\rm out}}{t_f} = \frac{1}{t_f} \left(\frac{k_B\Delta T}{2}\mathcal{A}_h - \frac{\gamma}{2t_f}\ell^2\right)
\end{equation*}
is maximized for $t_f = \frac{2\gamma \ell^2}{k_B \Delta T \mathcal{A}_h}$, so that power and efficiency (defined in \eqref{eq:def_eff}) become
\begin{equation}\label{eq:max-p}
    P_{\rm max} = \frac{(k_B\Delta T)^2 \mathcal{A}_h^2}{8\gamma\ell^2},\mbox{ and } \eta^* = \frac{1}{2}.
\end{equation}
These depend on $\mathcal{A}_h,\ell$ and apply to all cycles (traversed in constant speed on the Wasserstein manifold, as explained earlier). The efficiency at maximum power $\eta^*$ matches the universal linear-response bound of $1/2$ 
\cite{vandenBroeck,esposito}.

\section{Concluding remarks}
The present work builds on \cite{olga} that derived quantitative bounds on power and efficiency for a thermodynamic engine that is based on the Brownian gyrator. The salient feature is to capitalize on a temperature gradient that can produce a torque on mechanical degrees of freedom, and thereby allow extracting work from the heat baths that are coupled via these same degrees of freedom.

Thermodynamic states are seen as distributions on the Wasserstein manifold (distributions metrized by the $2$-Wasserstein metric of Optimal Mass Transport theory). Lengths being traversed in a thermodynamic cycle quantify dissipation while suitably weighted area quantifies work produced during the cycle. Our analysis echoes that in \cite{olga} where similar conclusions where drawn under the assumption of isentropic cycles. Our results are more general since fluctuation of the entropy of thermodynamic states as they traverse a cycle can be used judiciously to reduce dissipation. Specifically, we obtain increased maximal work output, tighter bounds on efficiency \eqref{eq:efficiency}, inherent improvement in speed limits \eqref{eq:sl} and explicit expressions for maximum power and for efficiency at maximum power \eqref{eq:max-p}.

Future work may focus on losses due to {\em housekeeping} entropy production, that has not been dealt with in the current work (see  \cite{olga_entropy,hatano}). A holistic picture of how temperature gradients can be used to generate work and how protocols may be designed to minimize entropy production should be of great interest when studying biological engines. The distinguishing feature of such real-world embodiments is that the capacity of heat baths or of chemical potentials is not inexhaustible, and thereby, total entropy production should be contained as much as possible (see \cite{richens}). Ultimately, it would be of great interest to compare theoretical results to experimental data that pertain to flagellar and other biological engines.\\


\section*{Appendix}\label{section:app2}
Symbolic computations to derive the expressions for $\mathcal Q_{\rm qs}$ and $\mathcal Q_{\rm diss}$ in \eqref{eq:Qall}, using (\ref{eq:defQ}-\ref{eq:eq2}, \ref{eq:eq3}), can be carried out using the Mathematica$^\copyright$ code that follows.

\begin{widetext}

\begin{lstlisting}[style=Mathematica]
 $\uptheta$= Symbol["$\uptheta$"]; r = Symbol["r"]; z = Symbol["z"]; t = Symbol["t"];
 T = DiagonalMatrix[{Symbol["T1"],Symbol["T2"]}];
 r1 = {{Cos[-$\uptheta$/2],Sin[-$\uptheta$/2]},{-Sin[-$\uptheta$/2],Cos[-$\uptheta$/2]}}; r2 = {{Cos[$\uptheta$/2],Sin[$\uptheta$/2]},{-Sin[$\uptheta$/2],Cos[$\uptheta$/2]}};
 dg = {{Exp[z/2+r],0},{0,Exp[z/2-r]}};  Sigma = r1.dg.r2; Xi = {{1,0},{0,-1}};  Om = {{0,1},{-1,0}};
 dr = Symbol["dr"]; dz = Symbol["dz"]; d$\uptheta$ = Symbol["d$\uptheta$"];
 dSigma = Simplify[-r1.dg.r2*dz/2+r1.dg.Xi.r2*dr+r1.(dg.Om-Om.dg).r2*d$\uptheta$/2];
 expTSigma = Simplify[MatrixExp[-t*(Sigma)].T.MatrixExp[-t*(Sigma)]];
 integral = Simplify[Integrate[expTSigma,{t,0,Infinity}]];
 expdSigma = Simplify[MatrixExp[-t*(Sigma)].dSigma.MatrixExp[-t*(Sigma)]];
 integral2 = Simplify[Integrate[expdSigma,{t,0,Infinity}]];
 dQ = Symbol["kB"]*Simplify[Tr[integral.dSigma]]; dW = -Symbol["$\upgamma$"]/2*Simplify[Tr[integral2.dSigma]];
\end{lstlisting}
\normalsize
\end{widetext}

\bibliography{apssamp}

\end{document}